\documentclass{article}

\usepackage{siunitx}
\usepackage{arxiv}                  % ArXiv template
\usepackage{newtxtext,newtxmath}    % Math in Times
\usepackage{hyperref}       % hyperlinks
\usepackage{float}      % [H] float specifier
\usepackage{cite}       % ordering in \cite{}
\usepackage[margin=2cm,tableposition=top]{caption}    % make captions 1cm narrower than text
\DeclareCaptionLabelFormat{andtable}{#1~#2  \&  \tablename~\thetable}

\usepackage{amsmath,amsthm,mathtools}
\usepackage{interval,xfrac,bm}

\intervalconfig{soft open fences}

\usepackage{mathrsfs}

\renewcommand{\vec}{\bm}

\usepackage{xcolor}

\usepackage{siunitx}
\DeclareSIUnit\molar{\textsc{m}}
\DeclareSIUnit\calorie{cal}

\usepackage[version=4]{mhchem}

\makeatletter
\let\@Re\Re
\let\@Im\Im
\let\@exp\exp
\let\Re\relax
\let\Im\relax
\let\exp\relax
\DeclarePairedDelimiterXPP\Re[1]{\@Re}{\{}{\}}{}{#1}
\DeclarePairedDelimiterXPP\Im[1]{\@Im}{\{}{\}}{}{#1}
\DeclarePairedDelimiterXPP\exp[1]{\@exp}{\{}{\}}{}{#1}
\makeatother

\newcommand*{\tpose}{^{\mkern-1.5mu\mathsf{T}}}
\newcommand*\inv{^{-1}}

\DeclarePairedDelimiter\norm{\lVert}{\rVert}
\newcommand*{\pnorm}[2]{\norm{#2}_{#1}}

\DeclarePairedDelimiterXPP\spn[1]{\operatorname{span}}{\{}{\}}{}{#1}

\DeclarePairedDelimiterX\inp[2]{\langle}{\rangle}{#1 , #2}

\DeclareMathOperator\compose\circ

\newcommand*\boundary\partial

\newcommand*\Naturals{\mathbb{N}}
\newcommand*\Reals{\mathbb{R}}

\newcommand*\directsum\oplus
\newcommand*\cartprod\times
\newcommand*\directprod\times

\newcommand\SetSymbol[1][]{%
    \nonscript\:#1\vert
    \allowbreak
    \nonscript\:
    \mathopen{}}
\newcommand\given\SetSymbol
\DeclarePairedDelimiterX\Set[1]\{\}{%
    \renewcommand\given{\SetSymbol[\delimsize]}
    #1}

\newcommand\distributed\sim

\DeclarePairedDelimiterXPP\Prob[1]{\mathrm{P}}{(}{%
    \renewcommand\given{\SetSymbol[\delimsize]}
    #1}{)}{}
\DeclarePairedDelimiterXPP\Expect[1]{\mathbb{E}}{[}{%
    \renewcommand\given{\SetSymbol[\delimsize]}
    #1}{]}{}

\newcommand*\dd{\mathrm{d}}

\newcommand*\diff[2][]{\frac{\dd#1}{\dd#2}}

\newcommand*\tdiff[2][]{\tfrac{\dd#1}{\dd#2}}

\makeatletter
\newcommand{\eval}{\@ifstar{\@evalar}{\@eval}}
\newcommand{\@evalar}[3]{\left. #1 \right|_{#2}^{#3}}
\newcommand{\@eval}[4][]{#2 #1|_{#3}^{#4}}
\makeatother

\usepackage[title]{appendix}

\renewcommand{\appendixname}{Appendix}

\mathtoolsset{showonlyrefs}

\graphicspath{{Figures/}}

\title{A Polyaddition Model for the Prebiotic Polymerization of RNA and RNA-like Polymers}

\author{
  Alex Spaeth$^*$\\
  University of California, Santa Cruz\\
  Santa Cruz, CA\\
  \texttt{atspaeth@ucsc.edu} \\
  \And
  Mason Hargrave\thanks{Both authors contributed equally.}\\
  The Rockefeller University\\
  New York, NY \\
  \texttt{mhargrave@rockefeller.edu}
}

\begin{document}

\maketitle

\begin{abstract}
Implicit in the RNA world hypothesis is that prebiotic RNA synthesis, despite occurring in an environment without biochemical catalysts, produced the long RNA polymers which are essential to the formation of life. In order to investigate the prebiotic formation of long RNA polymers, we consider a general solution of functionally identical monomer units that are capable of bonding to form linear polymers by a step-growth process. Under the assumptions that (1) the solution is well-mixed and (2) bonding/unbonding rates are independent of polymerization state, the concentration of each length of polymer follows the geometric Flory-Schulz distribution. We consider the rate dynamics that produce this equilibrium; connect the rate dynamics, Gibbs free energy of bond formation, and the bonding probability; solve the dynamics in closed form for the representative special case of a Flory-Schulz initial condition; and demonstrate the effects of imposing a maximum polymer length. Afterwards, we derive a lower bound on the error introduced by truncation and compare this lower bound to the actual error found in our simulation. Finally, we suggest methods to connect these theoretical predictions to experimental results.
\end{abstract}


\section{Introduction}
The RNA world hypothesis maintains that RNA molecules, being capable of both performing functions and storing information, were the first self-replicating molecules in the origin of life \cite{gilbert1986origin,neveu2013strong}. Deamer et~al.~\cite{deamer2019hydrothermal} have advanced a specific theory that outlines the importance of RNA to the origins of life. Of particular interest is the formation of long RNAs called ribozymes which are capable of catalysis \cite{kruger1982selfsplicing, fedor2005catalytic}. A variety of ribozymes have been designed and synthesized \cite{bartel1993isolation}, including some capable of catalyzing the reactions necessary for RNA replication \cite{johnston2001rnacatalyzed}, and some capable of replicating other ribozymes \cite{wochner2011ribozymecatalyzed, attwater2013inice}. In theory, collections of ribozymes may form autocatalytic sets, leading to self replication and evolution \cite{kauffman1993origins,lancet1994emergence,vasas2012evolution, hordijk2014conditions}. In order for a ribozyme to exist in the first place, non-enzymatic RNA synthesis must have occured. Extensive experiments have been conducted on non-enzymatic RNA polymerization in various settings, including lipid-assisted synthesis, templating, and chemical activation of the phosphate \cite{orgel2004prebiotic}. Additionally, the effect of wet-dry cycling on RNA polymerization has been studied in simulation~\cite{higgs2016effect,ross2016dry,hargrave2018computational} as well as experiment \cite{rajamani2008lipidassisted,dasilva2015saltpromoted,deguzman2014generation}.

RNA polymerization occurs through dehydration synthesis: the ribose unit of one nucleotide bonds with the phosphate unit of another, releasing a single water molecule. This is a classic example of a polycondensation process \cite{flory1953principles}; however, classical models of polycondensation are not appropriate for RNA polymerization because they were developed for chemical batch reactors where the reaction product is continuously evacuated to increase yield \cite[ch.~2]{gupta1987reaction}. Instead, since the essential processes of life take place in aqueous solution, the condensate is negligibly small in comparison to the solution as a whole. As a result, the external concentration of water is approximately constant, so RNA polymerization is more accurately modeled as a polyaddition process.

Consider an experiment initially consisting of a solution of monomers (e.g. nucleotides) capable of bonding with each other to form polymers. Each monomer can support two bonds, one on its left and one on its right, so that these monomers can link together to form linear polymers of an arbitrary length. A contiguous chain of $k$ monomer units will henceforth be referred to as a $k$-mer, including the monomer case where $k=1$. For the sake of visualization, one can imagine each monomer unit as a puzzle piece with a \ce{A} terminus and an \ce{B} terminus. It is important to note that no matter how long a polymer becomes, it always has precisely one unbound \ce{A} terminus and one unbound \ce{B} terminus.

We now make the assumption that the system is well-mixed in the sense that all reactants move and interact freely independent of mass, polymerization status, etc. Under these conditions, the polyaddition interaction between \ce{A} and \ce{B} termini is described by Hill-Langmuir protein-ligand reaction kinetics; that is, the two termini bind to each other with a reaction rate constant $k_+$, and bonded \ce{A-B} pairs separate from each other at a rate $k_-$. These are assumed to be independent of the configuration of the reacting monomer units; that is, the bonding rate $k_+$ does not depend on whether each \ce{A} and \ce{B} terminus is the endpoint of a long polymer or of a free monomer, nor is the unbonding rate $k_-$ affected by the position of the \ce{A-B} bond within a polymer. Under these conditions, the reactions affecting each bonding site take the following simple form:
\begin{equation}
    \ce{A + B <-->[$k_+$][$k_-$] AB} \label{eq:polyaddition}
\end{equation}

\section{Flory-Schulz Polymer Length Distribution}\label{sec:steadystate}
We have assumed that all binding sites behave identically; this implies that each site has the same (potentially time-varying) probability $p$ of being occupied at any given time. This fact, independent of bonding and un-bonding rates, leads very directly to a geometric distribution of polymer length \cite{flory1953principles}. Alternatively, Higgs~\cite{higgs2016effect} provides a proof of the Flory-Schulz distribution as an equilibrium state characterized entirely by bonding and un-bonding rates as opposed to bonding probabilities.

To see how our bonding probability assumption leads to a geometric distribution, we can perform the thought experiment of randomly selecting a $k$-mer of any length from the solution. Moving from left to right along the $k$-mer, the probability of a bond existing between two consecutive monomer units is $p$. In this way, we can view each $k$-mer as a sequence of Bernoulli trials, where the length of the $k$-mer is the number of trials up to and including the first failure. The result is by definition a geometric distribution with parameter $p$, so the probability mass function $\rho(k)$ over polymer length is given for positive $k$ by:
\begin{equation}
    \label{eq:length-pmf}
    \rho(k) = (1 - p) p^{k-1}
\end{equation}

From this probability distribution over polymer length, we would like to find the expected concentration of each $k$-mer as a function of our total concentration $\ce{[U]}$ of monomer units. If we define $n_*$ to be equal to the total concentration of reactants, including monomers and polymers of all lengths, the expected concentration $n(k)$ of $k$-mers is given by multiplication with \eqref{eq:length-pmf} as follows:
\begin{equation}
    \label{eq:expected-concentration}
    n(k) = n_* \rho(k) = n_* (1-p) p^{k-1}
\end{equation}

However, we would like to express this result in terms of $\ce{[U]}$ rather than $n_*$ because $n_*$ varies with time as bonds break and reform, whereas $\ce{[U]}$ is fixed in a closed system. We can find the value of $n_*$ using conservation of mass:
\begin{equation}
    \label{eq:nstar}
    \ce{[U]} = \sum_{k=1}^\infty k n(k)
    = n_* (1-p) \sum_{k=1}^\infty k p^{k-1}
    = n_* (1-p)\inv
    \quad\implies\quad
    n_* = (1 - p) \ce{[U]}
\end{equation}

The distribution of polymer lengths for any bonding probability $p$ is given by substituting the value of $n_*$ into \eqref{eq:expected-concentration}:
\begin{equation}
    \label{eq:floryschulz}
    n(k) = (1-p)^2 p^{k-1} \ce{[U]}
\end{equation}

\subsection{Steady-State Bonding Probability from Reaction Rates}
We consider a step-growth polymerization process described by \eqref{eq:polyaddition}, and assume that the total number of reactants is large enough for the law of mass action to apply. Under these conditions, if we introduce the equilibrium constant $\kappa = k_- / k_+$, the steady-state concentration of \ce{A-B} bonds \ce{[AB]} is given by the Hill-Langmuir equation:
\begin{equation}\label{eq:raw-hill-langmuir}
    \ce{[AB]} = \ce{[U]}\frac{\ce{[A]}}{\ce{[A]} + \kappa}
\end{equation}

Here $\ce{[A]}$ is always equal to the total reactant concentration $n_*$ because each monomer or polymer has exactly one unbound A and one unbound B terminus. Thus we can calculate the steady-state bonding probability $P_b$:
\begin{equation}\label{eq:bonding_probability_1}
    P_b = \frac{\ce{[AB]}}{\ce{[U]}} = \frac{n_*}{n_* + \kappa} = \frac{(1 - P_b) \ce{[U]}}{(1 - P_b) \ce{[U]} + \kappa}
\end{equation}

Rearranging to solve for $P_b$ gives a quadratic equation with two real roots for positive values of $\kappa$. One of these roots is greater than 1 and thus cannot correspond to a probability, so the other must be the solution. We introduce the reduced rate constant $\bar\kappa = \kappa / 2\ce{[U]}$ and solve to find a value of $P_b$ which can be substituted into \eqref{eq:floryschulz}:
\begin{equation}
\label{eq:Pb_function_of_a}
    P_b  = 1 + \bar\kappa - \sqrt{\bar\kappa(2 + \bar\kappa)}
\end{equation}

\subsection{Thermodynamics of Bonding}\label{sec:thermo}

We have described the bonding sites as a vast number of non-interacting systems which alternate stochastically between discrete states. This means that the steady-state probability of bonding can be described by Boltzmann statistics if we associate a Gibbs free energy $\Delta G_b$ with the bound state:
\begin{equation}
    \label{eq:boltzmann}
    P_b = \frac{e^{-\Delta G_b / RT}}{1 + e^{-\Delta G_b / RT}}
    \quad \iff \quad
    \Delta G_b = -RT\ln\frac{P_b}{1 - P_b}
\end{equation}

For this system, the equilibrium constant $\kappa$ must have units of concentration, meaning that the commonly-employed expression $\kappa = e^{\Delta G_b / RT}$ is dimensionally inconsistent. Solving \eqref{eq:bonding_probability_1} for $\kappa$, then substituting \eqref{eq:boltzmann}, we find:
\begin{equation}
    \label{eq:kappa-gibbs}
    \kappa = \frac{\ce{[U]} e^{\Delta G_b / RT}}{1 + e^{-\Delta G_b / RT}}
\end{equation}

\begin{figure}[tb]
    \centering
    \begin{minipage}{3in}
    \includegraphics{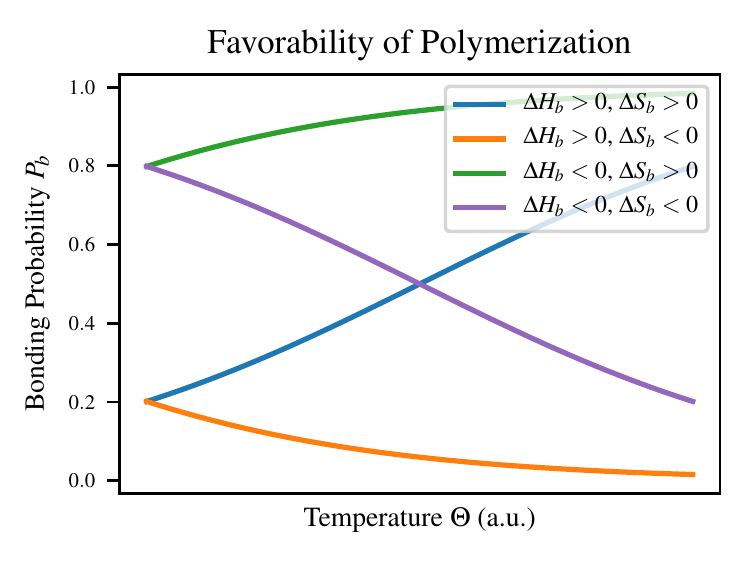}
    \end{minipage}
    \begin{minipage}{2.5in}
    \centering
    \begin{tabular}[t]{ccl}
        $\Delta H_b$ & $\Delta S_b$ & Effect on $P_b$ \\\hline
        + & + & $P_b > 0.5$ above $\frac{\Delta H_b}{\Delta S_b}$\\
        + & - & $P_b < 0.5$ at all $T$\\
        - & + & $P_b > 0.5$ at all $T$\\
        - & - & $P_b > 0.5$ below $\frac{\Delta H_b}{\Delta S_b}$\\
    \end{tabular}
    \end{minipage}
    \stepcounter{table}
    \captionsetup{labelformat=andtable}
    \caption{A comparison of the four different cases for the signs of $\Delta H_b$ and $\Delta S_b$. When $\Delta H_b$ and $\Delta S_b$ have the same sign, there is a critical temperature $T_c = \frac{\Delta H_b}{\Delta S_b}$ at which $\Delta G_b = 0$, so $P_b = 50\%$ and polymerization changes between being favorable and unfavorable. When the signs differ, however, polymerization is either favorable or unfavorable regardless of temperature.}
    \label{fig:thermodynamical}
\end{figure}

The relation defining the Gibbs free energy implies a functional dependence between $\Delta G_b$ and temperature: $\Delta G_b = \Delta H_b - T \Delta S_b$, where $\Delta H_b$ and $\Delta S_b$ are the enthalpy and entropy of bonding respectively. This means that depending on the signs of these two quantities, a polymerization reaction may change favorability depending on temperature \cite[table~1-3]{voet2011fundamentals}; both $P_b$ and $\kappa$ will vary with temperature to reflect this. The four cases are compared in \autoref{fig:thermodynamical}.

We expect intuitively that in most polymerization reactions $\Delta S_b$ would be negative due to the increased order, meaning that polymerization would have to be enthalpically favorable in order to be observed at all. This explains the observation that polymerization is favorable at low temperatures but polymers break down as temperature increases, for example in self-assembly of nanowires \cite{gao2019synthesis}.

  The suggestion that polymerization ought only to proceed when bond formation is enthalpically favorable may appear to conflict with the fact that the formation of an ester linkage between a sugar and a phosphate group is endergonic under standard conditions \cite[table~13-4]{nelson2013lehninger}.
  However, this is a consequence of entropic unfavorability---by confining the reactants to the surface of a microdroplet, Nam et~al. were able to nearly eliminate the contribution of the $T\Delta S$ term to the free energy of esterification, revealing a favorable negative value of $\Delta H$ % = \SI{-0.9}{kcal/mol}$ 
  \cite{nam2017abiotic}. 
  Other means of reducing the entropic unfavorability of polymerization such as mineral surface adsorption \cite{orgel1998polymerization}, restriction to small cavities \cite{monnard2003eutectic}, or the excluded volume effect of crowding \cite{ellis2001macromolecular} can also increase the favorability of polymerization.

\section{Dynamics}\label{sec:analytical}
In this section, we look at another way of thinking about our chemical system. In particular, we consider a countably infinite family of reaction equations which describe the way in which $i$-mers and $j$-mers bond to form $(i+j)$-mers, represented with the chemical symbols $\ce{P_i}$, $\ce{P_j}$, and $\ce{P_{i+j}}$. The chemical equations in this family are of the form:
\begin{equation}
    \ce{P_i + P_j <-->[$k_+$][$k_-$] P_{i+j}} \label{eq:reaction_equation_kmers}
\end{equation}

It is perhaps not immediately obvious that \eqref{eq:reaction_equation_kmers} describes the same system as \eqref{eq:polyaddition}, but in fact they are two different views of the same chemical process. From the perspective of bond formation, a $k$-mer is identical to a monomer in that it has precisely one \ce{A} terminus and one \ce{B} terminus. In this view, \eqref{eq:reaction_equation_kmers} is derived from splitting up the single reaction equation \eqref{eq:polyaddition} into separate chemical equations describing the behavior of each possible configuration of \ce{A} and \ce{B} termini: the \ce{A} terminus is the end of an $i$-mer, and the \ce{B} terminus is the end of a $j$-mer.

\subsection{Continuous Dynamics}
\label{sec:dynamics}
We have found a set of chemical equations which describe the interactions of individual $k$-mers $\ce{P_k}$. This is fundamentally a stochastic jump process describing discrete numbers of $k$-mers, but in the thermodynamic limit as the number of reactants grows very large, we can concern ourselves with the deterministic, continuous evolution of the \emph{expected} concentration $n(k)$ of $k$-mers.

Our dynamics can be written as a system of differential equations describing the time derivative of $n(k)$. As is usual for deriving mass-action differential equations from systems of chemical equations, we find the time derivative of $n(k)$ by a summation over each place where \ce{P_k} occurs in the system of chemical equations: if it is on the left-hand side, a negative contribution is made to $\diff{t} n(k)$, and if on the right, the contribution is positive.

Any given $\ce{P_k}$ can appear in all three positions in the chemical equation \eqref{eq:reaction_equation_kmers}. For each equation where $\ce{P_k}$ appears as the first term on the left side (i.e. for each possible synthesis partner $j\in\Naturals$), we lose $\ce{P_k}$ at a rate $k_+\ce{[P_k][P_j]}$, but gain it at a rate $k_-\ce{[P_{k+j}]}$. Each of those contributions should also be doubled to handle the functionally identical case where $\ce{P_k}$ appears as the second term on the left side. Finally, when $\ce{P_k}$ appears on the right side, for each possible split point $j \in \Set{1 \ldots k-1}$, we gain $\ce{P_k}$ at a rate $k_+\ce{[P_{k-j}][P_j]}$ and lose it at a rate $k_-\ce{[P_k]}$. The facts above can be consolidated into a single differential equation describing the evolution of $n(k) = \ce{[P_k]}$ as follows:
\begin{equation} \label{eq:rate-dynamics}
    \diff[n(k)]{t} = \sum_{j=1}^\infty 2k_- n(k+j) - 2k_+ n(k) n(j)
    \quad + \quad
    \sum_{j=1}^{k-1} k_+ n(j) n(k-j) - k_- n(k)
\end{equation}

% The collection of these equations forms an infinite-dimensional dynamical system whose state vectors $n$ are sequences of expected concentrations, where the $k$th element of the sequence is written as $n(k)$. Physically meaningful state vectors are those which have well-defined mass and no negative components. We define ``mass'' in this context using the linear operator $M$ defined by $M n = \sum k n(k)$, which computes the total expected mass of our $k$-mer solution as a multiple of the molar mass of a single monomer unit. For concentration vectors with non-negative components, $M n \ge \pnorm{1}{n} \ge \pnorm{2}{n}$, so all physically meaningful state vectors are elements of the space $\ell^2$ of square-summable sequences.

\subsection{Reduction to One Dimension}
\label{ssec:reduction}
We consider the special case where the initial condition is a Flory-Schulz distribution \eqref{eq:floryschulz} with rate parameter $p(0)$. For example, the $p(0)=0$ case would be a solution consisting entirely of monomers, and is particularly relevant as it is a popular experimental initial condition \cite{deguzman2014generation,costanzo2009generation,rajamani2008lipidassisted, dasilva2015saltpromoted}.

The derivation of the Flory-Schulz distribution holds for all time in a well-mixed step-growth polymerization process, even as the distribution parameter $p$ evolves. This has been predicted theoretically \cite{flory1944thermodynamics} and demonstrated experimentally \cite{luo2017polymerizationlike,gao2019synthesis,yang2018supramolecular}. We would like to calculate the rate at which the distribution parameter $p$ changes with time. A generalization of this problem was discussed in the context of self-assembling nanoparticles by Gu et~al \cite[SI 2]{gu2019reversible}.

Applying the principle of mass action to \eqref{eq:polyaddition} to calculate the time derivative of the total concentration of bonds \ce{[AB]}, then dividing through by \ce{[U]} gives:
\begin{equation}\label{eq:bonds}
    \diff[p]{t} = \diff{t} \frac{\ce{[AB]}}{\ce{[U]}} = \ce{[U]}\inv k_+ (\ce{[U]} - \ce{[AB]})^2 - k_- \ce{[U]}\inv \ce{[AB]} = \ce{[U]} k_+ (1-p)^2 - k_- p
\end{equation}

\begin{figure}[tb]
    \centering
    \includegraphics{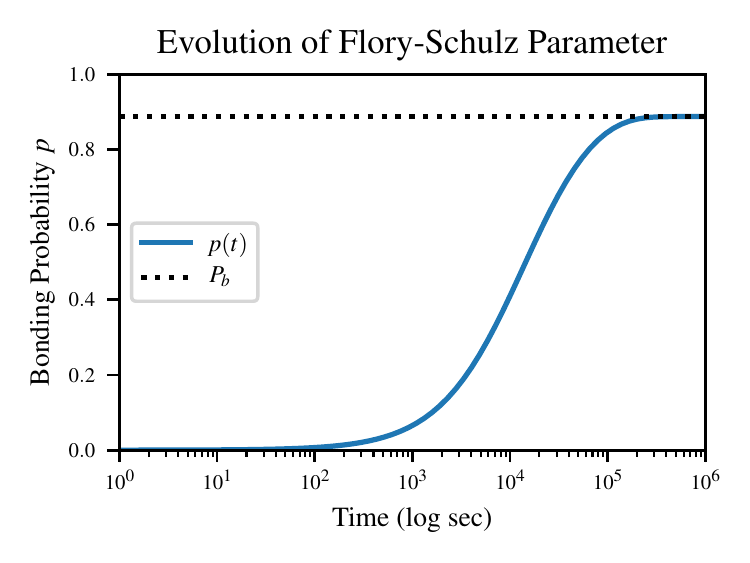}
    \includegraphics{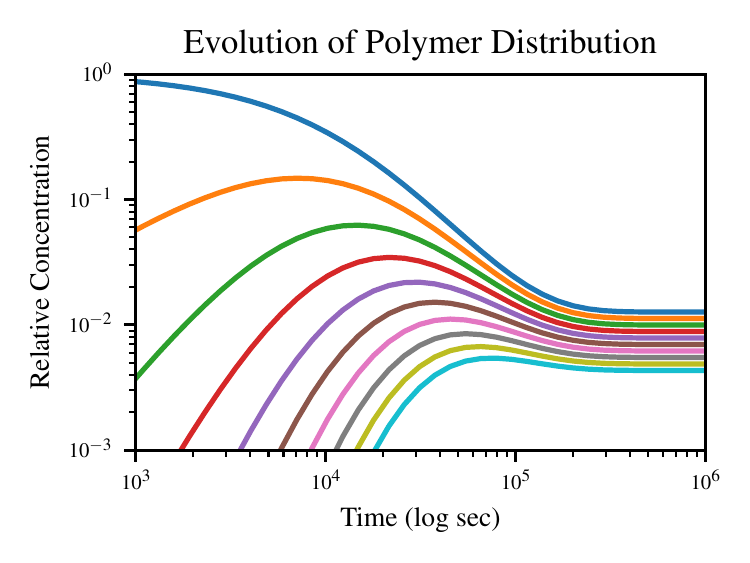}
    \caption{Closed-form solution to the dynamics of the Flory-Schulz rate parameter $p$ starting from an initial condition $p=0$, corresponding to an all-monomer solution. The parameter itself is shown on the left, and the resulting concentrations of $k$-mers for $k$ from 1 (blue) to 10 (cyan) are shown on the right.}
    \label{fig:analytical}
\end{figure}

\subsection{Closed-Form Solution}\label{ssec:closedform}
For the special case we just considered where the initial distribution is Flory-Schulz, the system has been reduced to the one-dimensional ODE \eqref{eq:bonds}. We can go one step further: this ODE is separable and admits a closed-form solution. In preparation for this, we will perform some simplifications. First, recall that the steady-state bonding probability $P_b = 1 + \bar\kappa - \Delta$, where $\bar\kappa = \frac{k_- / k_+}{2 \ce{[U]}}$ and $\Delta = \sqrt{\bar\kappa(2 + \bar\kappa)}$.  We nondimensionalize the ODE \eqref{eq:bonds} by setting $\tau = 2k_+ \ce{[U]} t$, transforming the equation into:
\begin{equation}
    \label{eq:dimensionless}
    \diff[p]{\tau} = \tfrac12 (1-p)^2 - \bar\kappa p
\end{equation}

The result is a separable ODE, allowing us to write:
\begin{equation}
    \int \frac{\dd p}{\frac12 (1-p)^2 - \bar\kappa p} = \int \dd \tau = \tau + c
\end{equation}

This gives us $\tau$ as a function of $p$, which can be inverted to give a solution to the ODE:
\begin{equation}
    p(\tau) = 1 + \bar\kappa - \Delta \tanh(\tfrac12 \Delta \tau + c)
\end{equation}

We can fix $p(0)$ to solve for the value of $c$:
\begin{equation}
    c = \tanh\inv\left(\tfrac{1 + \bar\kappa - p(0)}{\Delta}\right)
\end{equation}

Finally, we can recover the original time parameterization by replacing $\tau$ with $2 k_+ \ce{[U]} t$, which gives the parameter of the Flory-Schulz distribution as a function of time:
\begin{equation}
    \label{eq:solution}
    p(t) = 1 + \bar\kappa - \Delta \tanh\left(\Delta k_+ \ce{[U]}  t + \tanh\inv\left(\tfrac{1 + \bar\kappa - p(0)}{\Delta}\right)\right)
\end{equation}

As $\tau \to \infty$, the tanh function asymptotically approaches a value of 1 regardless of initial condition, which recovers the previously derived steady-state value $P_b$. \noeqref{eq:solution}

\section{Numerical Treatments} \label{sec:num}
We have derived the rate dynamics of interacting $k$-mers \eqref{eq:rate-dynamics} from the family of reaction equations \eqref{eq:reaction_equation_kmers}. However, because the state vectors lie in an infinite-dimensional space, physically realizable numerical methods require us to approximate these dynamics in finitely many dimensions.
From our work in \autoref{sec:steadystate}, we know that the expected number of extremely long polymers tends to be low due to the geometrically-distributed equilibrium state. Therefore, we can achieve very low error by introducing a constraint $d$ on the maximum length of polymers to be considered. This effectively constrains the system from the infinite-dimensional space $\ell^2$ down to the finite-dimensional $\Reals^d$.

  \subsection{Choice of Parameter Values}
  An experimentalist investigating polymerization in the lab might choose a set of representative conditions in the form of an initial temperature, pH, total concentration of monomer units, and presence of other cofactors such as salts.  The dynamics of the system are determined by the rate constants $k_+$ and $k_-$, which are a function of the experimental conditions; although the effect of pH and salt cofactors on hydrolysis have been studied in depth, e.g. by Oivanen et~al.~\cite{oivanen1998kinetics}, the effects of the same on synthesis rates have remained obscure.

  In our setting, we consider conditions common to experiments investigating the hot-spring origins of life hypothesis \cite{rajamani2008lipidassisted,deguzman2014generation,ross2016dry}, with temperature $T = \SI{85}{\celsius}$ and pH of approximately 3.  This allows us to take the approximate value of $k_-$ from the experimental results of Oivanen et~al.~\cite{oivanen1998kinetics} for similar conditions.  As noted in \autoref{sec:thermo}, although ester formation is endergonic under standard conditions, it is enthalpically favorable and can be made spontaneous by decreasing its entropic unfavorability. We assume that $\Delta G_b$ has been brought down by some means to an illustrative negative value, and compute the corresponding value of $k_+$.

\begin{table}
    \centering
    \caption{Parameter values used in all numerical simulations.} % The computed steady-state bonding probability $P_b$ is not used in simulations except to calculate asymptotic error.}
    \begin{tabular}{clc}
         Parameter & Description & Value  \\\hline
         $\ce{[U]}$ & initial monomer concentration & \SI{1}{\molar}\\
         $\Delta G$ & Gibbs free energy of bonding & \SI{-1.5}{kcal/mol}\\
         $k_-$ & unbonding rate & \SI{e-6}{s^{-1}}\\
         $k_+$ & bonding rate constant & \SI{7.4e-5}{s^{-1} mol^{-1}}\\
         $P_b$ & steady-state bonding probability & $89\%$
    \end{tabular}
    \label{tab:parameters}
\end{table}

\subsection{Truncation}
Although we seek to truncate the system to a finite dimension $d$, we do this not by throwing away polymers which become too large, but rather by eliminating the formation of longer polymers in the first place. This means that we approximate the family of reaction equations \eqref{eq:reaction_equation_kmers} by prohibiting all reactions which include a reactant of length greater than $d$:
\begin{equation}\label{eq:reaction_equation_truncated}
    \ce{P_i + P_j <-->[$k_+$][$k_-$] P_{i+j}} \quad\text{for }i+j \le d
\end{equation}

The dynamics can be derived from the reaction family \eqref{eq:reaction_equation_truncated} in exactly the same way that \eqref{eq:rate-dynamics} was derived from \eqref{eq:reaction_equation_kmers}, the only difference being that the first sum becomes finite due to the truncation. The resulting system of ODEs, describing the evolution of a state vector $\vec x \in \Reals^d$ whose components $x_k$ represent the concentration of $k$-mers, is given by:
\begin{equation}\label{eq:truncated}
    \diff[x_k]{t} = \sum_{l=1}^{d-k} 2k_- x_{k+l} - 2k_+ x_k x_l
    \ + \
    \sum_{l=1}^{k-1} k_+ x_l x_{k-l} - k_- x_k
\end{equation}

A perhaps more obvious method of truncation would be to keep the exact original form of \eqref{eq:rate-dynamics}, but ignore lengths above $d$ by taking $n(k) = 0$ for $k > d$. However, this approach leads to unsatisfactory results because it is equivalent to permanently deleting any $k$-mer which forms with $k>d$. Since the mass associated with these deleted $k$-mers is never returned to the system, mass is continually being lost, so the system asymptotically approaches a steady state at $\vec x = \vec 0$.

\subsection{Simulations}
\begin{figure}[tb]
    \centering
    \includegraphics{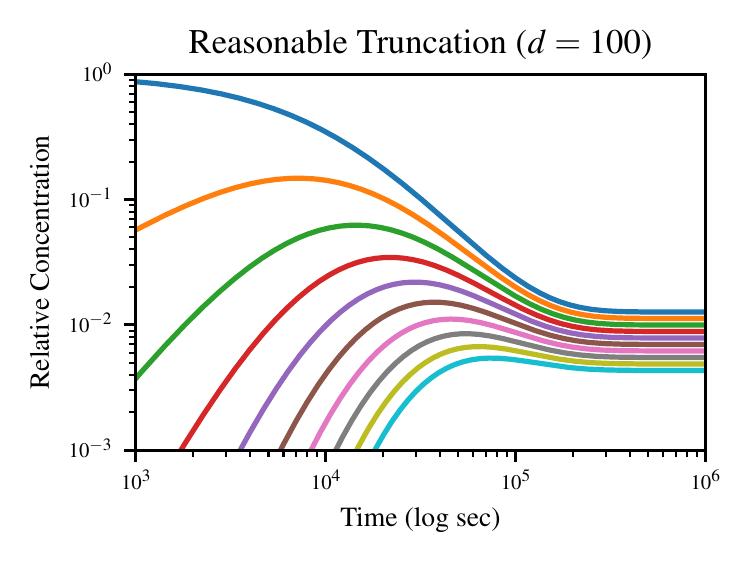}
    \includegraphics{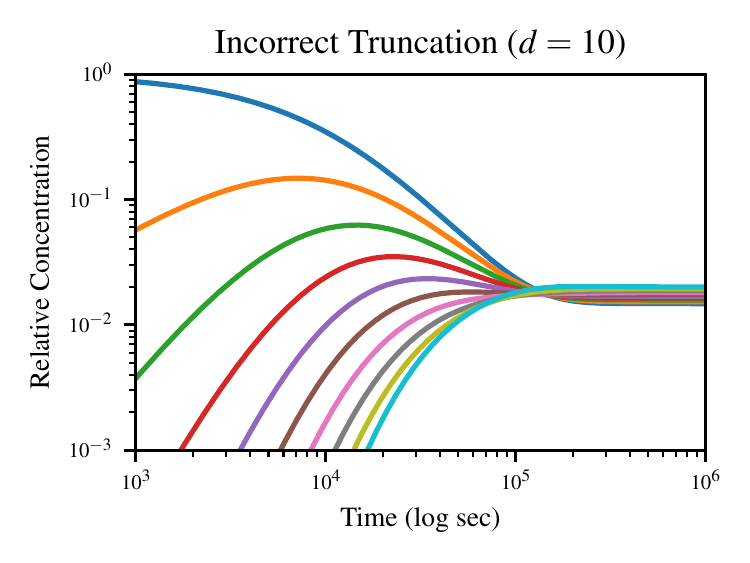}
    \caption{Concentration of $k$-mers for $k$ from 1 (blue) to 10 (cyan), for truncation lengths $d=100$ (left) and $d=10$ (right). The $d=100$ case, visually identical to the results shown in \autoref{fig:analytical}, reaches the correct geometric distribution, whereas the $d=10$ case goes through a nonphysical inversion near time $t=10^5$.}
    \label{fig:dynamics}
\end{figure}

To demonstrate the dynamics of the system and the effects of truncation, we numerically solve \eqref{eq:truncated} starting from an initial solution of exclusively monomers for the truncation lengths $d=100$ and $d=10$, and plot the concentration of $k$-mers up to length 10 over time in \autoref{fig:dynamics}. All simulations were run using the DifferentialEquations.jl package \cite{rackauckas2017differentialequations} with parameters other than $d$ held fixed; these values are given in \autoref{tab:parameters}.

The expected equilibrium state is the geometric distribution \eqref{eq:floryschulz}, which would appear uniformly spaced on a logarithmic plot, with the dimer concentration equal to $P_b$ multiplied by the monomer concentration and so on. In the case where $d=100$, this is exactly what we observe; however, when we truncate to $d=10$, the distribution goes through an inversion after which $d$-mers, rather than monomers, dominate. Since truncation depends on the assumption that longer polymers are negligible, this is obviously nonphysical.

\begin{figure}[tb]
    \centering
    \includegraphics{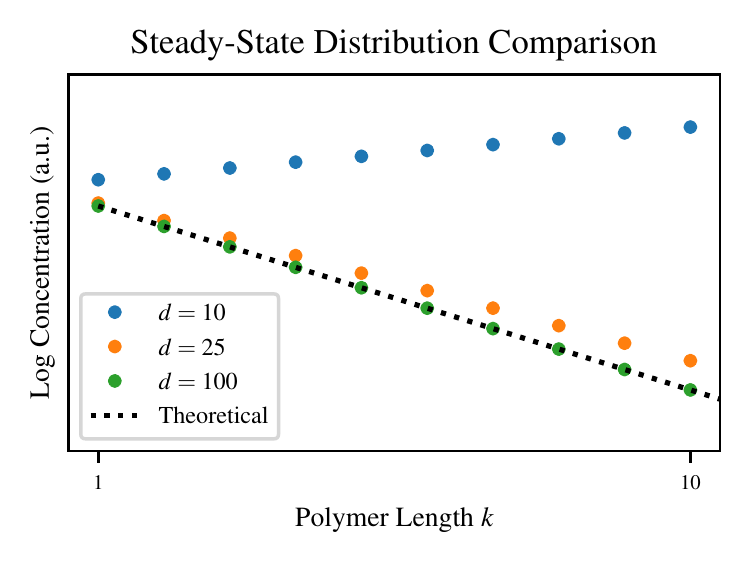}
    \caption{The steady state concentration distribution for $d=10$, $d=25$, and $d=100$ compared to the closed-form solution. By $d=100$, the numerical and analytical solutions are indistinguishable.}
    \label{fig:distributions}
\end{figure}

Although each of our simulations converges to some steady-state distribution, the degree of agreement with our theoretical prediction varies depending on the truncation length $d$. To visualize this, \autoref{fig:distributions} plots the steady state distributions for three values of $d$ compared to the theoretical steady-state Flory-Schulz distribution.

\subsection{Error Bound}
\label{ssec:bounds}
Since the Flory-Schulz distribution of polymer length which is the solution to the system of reaction equations \eqref{eq:rate-dynamics} includes a non-zero expected concentration for polymers longer than any finite $d$, it is impossible for the truncated probability distribution which is the solution to \eqref{eq:truncated} to be identical to the infinite-dimensional solution. As noted above, the dynamics of \eqref{eq:truncated} are exactly the result of constraining the dynamics of \eqref{eq:rate-dynamics} to finite maximum polymer length while preserving conservation of mass. Therefore, the distance between the true solution $n(k)$ and its projection $\hat n(k)$ onto the set of $d$-dimensional distributions with the correct total mass provides a lower bound to the error of \emph{any} mass-preserving truncation of the reaction family \eqref{eq:reaction_equation_kmers}. We can use the ``mass operator'' $M n = \sum k n(k)$, which counts the total concentration of monomer units in the system, to write this projection as:
\begin{equation}\label{eq:projection}
    \underset{\hat n \in \ell^2}{\operatorname{minimize}}\ \pnorm{2}{\hat n - n} \quad\text{ subject to } M \hat n = \ce{[U]} \text{ and } \hat n(k) = 0 \ \forall k > d
\end{equation}

% We can eliminate the second constraint and reduce the problem to finitely many dimensions by observing that if we introduce $x$ and $\hat x$ as $d$-dimensional truncations of $n$ and $\hat n$ and $y = n-x$ the residual of the former, we can apply the fact that $\hat n(k) = 0$ for $k > d$ to find:
  We can simplify the objective by separating the portion of $\hat n$ which is allowed to vary from the infinite ``tail'' which is fixed to zero. Introducing the $d$-dimensional truncations $x$ and $\hat x$ of $n$ and $\hat n$ respectively, and letting $y = n - x$ to capture the error in the tail, we have:
\begin{equation}\label{eq:normsquared}
    \pnorm{2}{\hat n - n}^2
    = \sum_{k=1}^\infty (\hat n(k) - n(k))^2
    = \sum_{k=1}^d (\hat n(k) - n(k))^2 + \sum_{k=d+1}^\infty n(k)^2
    = \pnorm2{\hat x - x}^2 + \pnorm2{y}^2
\end{equation}

Now we can change variables to $\delta x = \hat x - x$ and find the optimal projection using ordinary least squares. In plainer language, the problem being solved is to find the smallest correction $\delta x$ whose total mass is equal to that of the missing ``tail'' $y$. The finite-dimensional version of the mass operator $M_d x = \sum_{k=1}^d k x_k$ can be constructed as a $1\times d$ matrix whose entries are ascending integers.

\begin{equation}\label{eq:leastsquares}
    \underset{\delta x \in \Reals^d}{\operatorname{minimize}}\ \pnorm{2}{\delta x}^2 \quad\text{ subject to } M_d (x + \delta x) = \ce{[U]} \ \implies\ M_d \delta x = \ce{[U]} - M_d x = M y
\end{equation}

The well-known closed-form solution to the ordinary least squares problem \eqref{eq:leastsquares} is:
\begin{equation}
    \delta x = (M_d M_d\tpose)\inv M_d\tpose M y
\end{equation}

This can be brought into more elementary terms by calculating:
\begin{align*}
    & M y = \sum_{k=d+1}^\infty k n(k) = \ce{[U]}(1 - p)^2 \sum_{k=d+1}^\infty k p^{k-1} = \ce{[U]}(1 + d(1-p)) p^d\\
    & \pnorm{2}{y}^2 = \sum_{k=d+1}^\infty n(k)^2 = \sum_{k=d+1}^\infty (1-p)^4 \ce{[U]}^2 p^{2(k-1)} = \frac{(1-p)^4 \ce{[U]}^2 p^{2d}}{1 - p^2}\\
    & M_d M_d\tpose = \sum_{k=1}^d k^2 = \tfrac16 d (d+1) (2d+1)
\end{align*}

and
\begin{equation}
    \pnorm2{\delta x}^2 = \delta x\tpose \delta x =  My M_d (M_dM_d\tpose)^{-1} (M_dM_d\tpose)^{-1} M_d\tpose My
    = \frac{(My)^2}{M_dM_d\tpose}
\end{equation}

We can now directly calculate the total distance of the projection using \eqref{eq:normsquared}.
\begin{equation}
    \pnorm{2}{\hat n - n}^2 = \pnorm2{\delta x}^2 + \pnorm2{y}^2 = \frac{6\ce{[U]}^2(1 + d(1-p))^2 p^{2d}}{d(d+1)(2d+1)} + \frac{(1-p)^4 \ce{[U]}^2 p^{2d}}{1 - p^2}
\end{equation}

This provides a lower bound on the sum-squared error of any solution with maximum polymer length $d$ and the same total mass as the previously proven solution $n(k)$. In order to make this result more directly comparable between different parameter values, however, we will use the relative error:
\begin{equation}\label{eq:errorbound}
    E(p) = \sqrt{\frac{\pnorm2{\hat n - n}^2}{\pnorm2{n}^2}} = p^d\sqrt{1 + \frac{6(1 + d(1-p))^2(1-p^2)}{d(d+1)(2d+1)(1-p)^4}} > p^d
\end{equation}

\subsection{Applying the Error Bound}
This result $E$, which is strictly greater than but asymptotically equal to $p^d$, provides an absolute lower bound on the $\ell^2$ error between the instantaneous distribution of polymer lengths and any mass-conserving finite approximation to this distribution. In other words, error terms on the order of $p^d$ arise in any simulation of our chemical system, so long as the simulation (a) produces finite-dimensional results and (b) obeys conservation of mass. These errors can be surprisingly large even for quite reasonable-sounding $d$ as $p$ approaches 1. This result is relatively insensitive to the choice of error metric; although we specifically investigate the case of $\ell^2$ norm, other metrics which we tested in simulation also produced error on the order of $p^d$.

\begin{figure}[tb]
    \centering
    \includegraphics{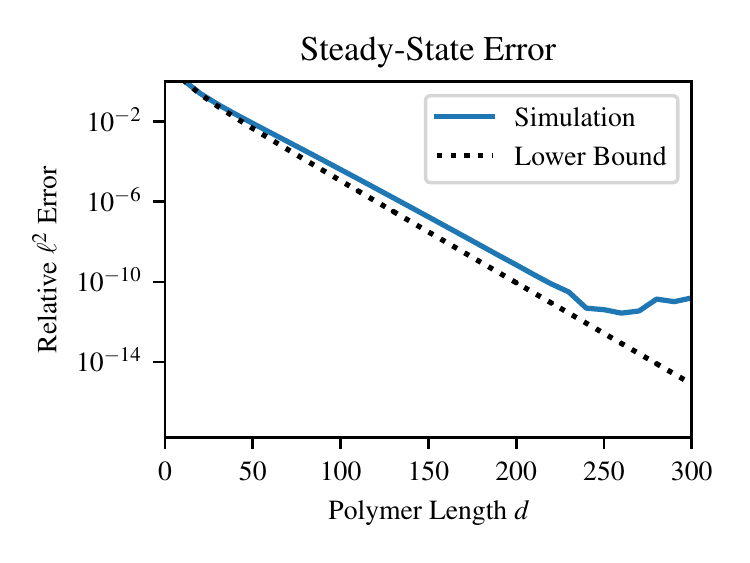}
    \includegraphics{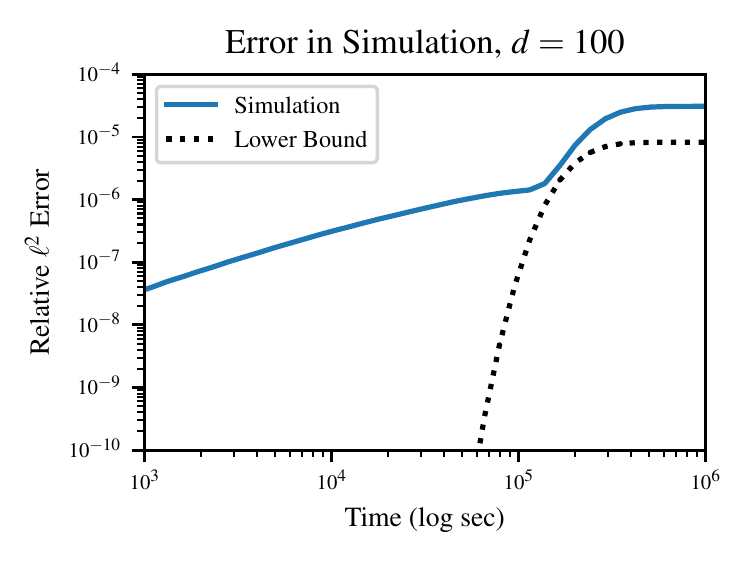}
    \caption{Comparison between the error bound \eqref{eq:errorbound} and the actual error in the results of our simulation. The left figure depicts the steady-state error and the theoretical lower bound $E(P_b)$ as a function of $d$, and on the right is the time evolution of the error in a single simulation for $d=100$, compared to the bound $E(p)$ computed from the instantaneous analytical value of $p$ as in \autoref{fig:analytical}.}
    \label{fig:errors}
\end{figure}

It is important to emphasize that this is a lower bound on error; this does not guarantee that a certain choice of $d$ will produce less than a certain error, which is in fact impossible without being more specific about the method of solution. For example, in \autoref{fig:errors}, above a certain truncation length of about $d=250$, the finite precision of the solver becomes more of a limiting factor than the truncation error. Likewise, early in the dynamical simulation when the instantaneous value of $p$ is very small, the error bound is practically useless. Instead, this bound guarantees that any simulation which chooses $d$ too small will produce at least a certain specified error.

As an example of applying this error bound in practice, if we consider a system with $\Delta G_b = \SI{-3.5}{kcal/mol}$, corresponding to a steady-state bonding probability $P_b = (1 + e^{\Delta G_b / RT})\inv \approx 99.3\%$, we can numerically solve \eqref{eq:errorbound} for $d$ to find that a simulation with $d < 700$ cannot have final relative error less than $1\%$.
The simpler error bound $E > P_b^d$ is even easier to apply: any truncation length $d < \log_{P_b} E^*$ must produce final relative error $E > E^*$. In the above case, this laxer bound is only able to rule out truncations up to $d = 632$, but the ease of calculation makes this bound probably more useful than the tighter one.

\section{Comparison to Experiment}
  The main value of a theoretical model of any physical process is the predictive and interpretive power it brings to designing and analyzing experimental results. In this section, we present a few commonly measured experimental quantities and the predictions our theory makes about them.

\subsection{Critical Concentration}
  
  As the total concentration \ce{[U]} of monomer units increases, the concentration of monomers in the final solution approaches a fixed value $C_c$ called the critical concentration \cite[sec.~3.2]{oosawa1975thermodynamics}. This value is a function only of the rate constants, and can be calculated by taking the limit of $n(1)$ as \ce{[U]} is taken to infinity. The monomer concentration $n(1)$ is given by substituting \eqref{eq:Pb_function_of_a} into the Flory-Schulz distribution \eqref{eq:floryschulz} with $k=1$:
  \begin{equation}
    n(1) = (1 - P_b)^2 \ce{[U]} = \left(\bar\kappa^2 - 2\bar\kappa\sqrt{\bar\kappa(2 + \bar\kappa)} + \bar\kappa(2 + \bar\kappa)\right) \ce{[U]}
  \end{equation}

  As \ce{[U]} grows to infinity, the reduced equilibrium constant $\bar\kappa = \kappa/2\ce{[U]}$ goes to zero, so we can calculate $C_c$ as:
  \begin{equation}\label{eq:critical-concentration}
    C_c = \lim_{\ce{[U]} \to \infty} n(1) = \lim_{\ce{[U]} \to \infty} 2\bar\kappa\ce{[U]} + o(\bar\kappa^2) = \kappa
  \end{equation}
  
  The critical concentration provides an alternate route to experimental determination of rate constants; since it is likely easier to measure the monomer concentration than to find the complete length distribution, it is possible to use the critical concentration to find one rate constant given the other.

\subsection{Polymer Yield}
Experimental studies commonly report the polymer yield, the fraction of mass which is converted to polymers at equilibrium. We can compute this mass conversion efficiency $\eta$ as:
\begin{equation}\label{eq:etam}
    \eta = \frac{\ce{[U]} - n(1)}{\ce{[U]}} = 1 - (1 - P_b)^2 = P_b (2 - P_b)
\end{equation}

The analogous quantity derived from concentrations rather than masses is simply equal to $P_b$, since the monomer concentration ratio is exactly $1-P_b$.

\subsection{Mass Distribution}
It is frequently easier to measure the \emph{mass} rather than the concentration of polymers of each given length. For example, the output of high-performance liquid chromatography (HPLC) is a ``spectrum'' where the height and location of each peak corresponds approximately to the total and per-molecule mass of a reaction product respectively. For comparison with such results, we follow Flory's treatment of the polymer mass distribution $m(k)$, which can be defined in terms of the concentration distribution $n(k)$ \eqref{eq:expected-concentration} as follows \cite{flory1936molecular}:
\begin{equation}
    m(k) = k m_0 n(k)
\end{equation}

Here, each term $m(k)$ is the theoretical total mass of $k$-mers, and $m_0$ is the molar mass of the corresponding monomer. Since $m(k)$ is a single peak distribution we can analyze an experimentally determined mass distribution $\hat{m}(k)$ by matching the position $k_*$ of the peak of $m(k)$ to the position $\hat{k}_*$ of the peak of $\hat{m}(k)$. We find the peak of our theoretical distribution by differentiating $m(k)$:
\begin{equation}
    \diff[m(k)]{k} = m_0 (1-p)^2 \ce{[U]} p^k (1 + k \ln p)
\end{equation}

The unique zero of this equation is $k_*$, so according to our theory, the mode of the mass distribution is related to the bonding probability by: 
\begin{equation}\label{eq:massmode}
    k_* = -1 / \ln p  
    \quad\iff\quad 
    p = e^{-1/k_*}
\end{equation}

In other words, given an experimental Flory-Schultz-like mass distribution, one can derive the value for the bonding probability $p$. At equilibrium, $p \to P_b$, so the Gibbs free energy of bond formation can be calculated using \eqref{eq:boltzmann}.

As a simple example of how these results may be applied, Monnard et~al. used HPLC to measure nonenzymatic polymerization of activated nucleotides \cite{monnard2003eutectic}. When multiple bases were mixed, the HPLC results are difficult to interpret, but in the case with pure uridine, the result appears to be a Flory-Schulz mass distribution with a mode of about 2. This corresponds by \eqref{eq:massmode} to $P_b \approx 60\%$. Substituting this value into \eqref{eq:etam}, we expect polymer yield of $\eta \approx 85\%$, consistent with their reported value of 88\%.

We can also estimate the free energy of bond formation under their experimental conditions in this way: when $P_b \approx 60\%$, the Boltzmann statistics of \eqref{eq:boltzmann} give a free energy $\Delta G \approx \SI{-0.2}{kcal/mol}$, corresponding to a process where bond formation is slightly favorable. Additionally, given the free energy at two different temperatures, we can separate the entropic and enthalpic contributions. Since this experiment was carried out at \SI{-18}{\celsius}, we can make a first approximation by assuming that our calculated value is directly comparable to the value for sugar-phosphate esterification under standard conditions of \SI{3.3}{kcal/mol} \cite[table~13-4]{nelson2013lehninger}.

The temperature of this experiment was \SI{43}{K} colder than the standard temperature of \SI{25}{\celsius}, and the concomitant free energy change was \SI{+3.5}{kcal/mol}. Since $\Delta G = \Delta H - T \Delta S$, a bit of algebra gives $\Delta S \approx -\SI{0.1}{\kilo\calorie\per\mole\per\kelvin}$ and $\Delta H \approx -\SI{20}{kcal/mol}$. These conditions correspond to the case in \autoref{fig:thermodynamical} where polymerization is favorable below a critical temperature $T_c \approx \SI{-15}{\celsius}$. It seems suspect, however, that $\Delta H$ would be so large \cite{nam2017abiotic}, suggesting that the role of entropy in bond formation is reduced by the environment of the eutectic ice-water mixture, in agreement with the commentary of Monnard et al. \cite{monnard2003eutectic}.

\subsection{Degree of Polymerization}
Another quantity commonly measured in experiments is the average degree of polymerization in obtained solutions \cite{costanzo2009generation}. Our model allows the evolution of this parameter to be calculated easily; by Carother's equation, the number average degree of polymerization is given by $\bar X_n = \frac{1}{1-p}$, so we can use \eqref{eq:bonds} and the chain rule to calculate:
\begin{equation}
    \diff[\bar X_n]{t} = \diff{p}\frac{1}{1-p} \diff[p]{t} = 2 k_+ \ce{[U]} - k_- \bar X_n (X_n - 1)
\end{equation}

This correctly simplifies in the irreversible case $k_-=0$ or when the degree of polymerization $\bar X_n = 1$ to a linear increase in degree of polymerization $\eval{\tdiff{t} \bar X_n}{\bar X_n=1}{} = 2 k_+ \ce{[U]}$. This linear dependence on initial concentration is in contrast to the quadratic dependence predicted for polycondensation. We are not aware of a study of the time course of $\bar X_n$ in RNA polymerization, but these dynamics have been observed in the synthesis of nanowires by Gao et.~al~\cite{gao2019synthesis}, in supramolecular polymerization of micelles by Lu et.~al~\cite{lu2019supramolecular}, and in simulations of interfacial polymerization by Xing et.~al~\cite{xing2019indepth}. In all three cases, the reversibility of the polymerization process is also in evidence due to the slowdown in the rate of increase in degree of polymerization.

\section{Discussion}
Our model provides an explicit description of the formation of RNA polymers in aqueous prebiotic conditions as is necessary for the RNA-world hypothesis. The mathematical and computational models presented in this paper generalize to all polymers that grow by polyaddition in well-mixed solutions. In these cases, as well as in polycondensation processes where the concentration of the condensate remains approximately constant, our models describe the dynamics of the length distribution as well as the eventual steady state. The time evolution of an initial Flory-Schulz distribution is completely determined by the evolution of the bonding probability, for which we have stated a closed form solution depending only on the forward and back reaction rates and the number of monomer units.

 In the simple case of an initial population of RNA monomers in absence of any cofactors, whenever polymerization is favorable in the sense that the Gibbs energy of bond formation is negative, the steady-state concentration of polymers is expected to exceed the concentration of monomers. However, we do not expect to see a population length inversion, sometimes dubbed a ``kinetic trap,'' in which polymers of certain lengths achieve a greater concentration than any shorter polymer. Any experimental deviation from these predictions with or without the presence of cofactors indicates the presence of significant higher-order effects (e.g. hairpin structures, cyclical polymers, catalysis) and may suggest future directions for mathematical models. The inclusion of the shielding of bonding sites from hydrolysis by virtue of the secondary structure of RNA in the model may increase the lifetime of long polymers in solution, leading to a recovery of kinetic trap-like behavior. Similarly, should the RNA in question be encapsulated in lipid vesicles small enough to introduce finite-size effects, or assisted in polymerization by association with a surface, then the statistics of our model may no longer apply.

Additionally, when considering the hydrothermal origins of life hypothesis advanced by Deamer et~al. it becomes important to consider the effects of wet-dry cycling on RNA polymerization \cite{deamer2019hydrothermal}. In the dry phase, polymerization is favorable due to the lack of hydrolysis but the well-mixed assumption is violated. In the wet phase, polymerization is unfavorable due to the presence of hydrolysis but the well-mixed assumption is upheld. Intuitively, this means that alternation between a relatively long dry phase and short wet phase allows for the ``well-enough mixing'' of the RNA molecules, such that the solution approaches the polymer distribution predicted for a dry phase with mixing \cite{higgs2016effect}. In short, wet-dry cycling leads to an effective increase in the probability of bonding and therefore increases the concentration of long polymers. Besides wet-dry cycling, another important feature of the hydrothermal hypothesis is the suggestion that biogenesis occurs at a certain optimal temperature: it is well-known that elevated temperature is required to increase the rate of biological reactions \cite{deamer2019hydrothermal}, but as temperature increases, polymerization becomes entropically unfavorable, suggesting a Goldilocks effect with regard to the ambient temperature at the origins of life.

\section{Supplementary Materials}
The Julia code used to perform these simulations is available at the following GitHub repository: \url{https://github.com/atspaeth/PolyadditionModel}.

\section{Funding} This research received no external funding.

\section{Acknowledgements}
We thank Dr.~David Deamer and Dr.~Bruce Damer at UCSC for the insightful conversations which catalyzed the generation of this model. Additionally, we express our great and sincere gratitude to Katerina Tadenev and Alexander Epstein for assistance with revising the manuscripts and advice concerning the interdisciplinary interpretability of the paper. Finally, we thank the anonymous reviewers for their inputs, which greatly improved this manuscript.

\bibliographystyle{unsrt}
\bibliography{references.bib}

\newcommand{\noopsort}[1]{}
\begin{thebibliography}{10}

\bibitem{gilbert1986origin}
Walter Gilbert.
\newblock Origin of life: {{The RNA}} world.
\newblock {\em Nature}, 319(6055):618--618, 1986.

\bibitem{neveu2013strong}
Marc Neveu, Hyo-Joong Kim, and Steven~A. Benner.
\newblock The ``strong'' {{RNA}} world hypothesis: {{Fifty}} years old.
\newblock {\em Astrobiology}, 13(4):391--403, 2013.

\bibitem{deamer2019hydrothermal}
David Deamer, Bruce Damer, and Vladimir Kompanichenko.
\newblock Hydrothermal chemistry and the origin of cellular life.
\newblock {\em Astrobiology}, 19(12), 2019.

\bibitem{kruger1982selfsplicing}
Kelly Kruger, Paula~J. Grabowski, Arthur~J. Zaug, Julie Sands, Daniel~E.
  Gottschling, and Thomas~R. Cech.
\newblock Self-splicing {{RNA}}: {{Autoexcision}} and autocyclization of the
  ribosomal {{RNA}} intervening sequence of tetrahymena.
\newblock {\em Cell}, 31(1):147--157, 1982.

\bibitem{fedor2005catalytic}
Martha~J. Fedor and James~R. Williamson.
\newblock The catalytic diversity of {{RNAs}}.
\newblock {\em Nature Reviews Molecular Cell Biology}, 6(5):399--412, 2005.

\bibitem{bartel1993isolation}
D.~Bartel and J.~Szostak.
\newblock Isolation of new ribozymes from a large pool of random sequences.
\newblock {\em Science}, 261(5127):1411--1418, 1993.

\bibitem{johnston2001rnacatalyzed}
Wendy~K. Johnston, Peter~J. Unrau, Michael~S. Lawrence, Margaret~E. Glasner,
  and David~P. Bartel.
\newblock {{RNA}}-catalyzed {{RNA}} polymerization: Accurate and general
  {{RNA}}-templated primer extension.
\newblock {\em Science}, 292(5520):1319--1325, 2001.

\bibitem{wochner2011ribozymecatalyzed}
Aniela Wochner, James Attwater, Alan Coulson, and Philipp Holliger.
\newblock Ribozyme-catalyzed transcription of an active ribozyme.
\newblock {\em Science}, 332(6026):209--212, 2011.

\bibitem{attwater2013inice}
James Attwater, Aniela Wochner, and Philipp Holliger.
\newblock In-ice evolution of {{RNA}} polymerase ribozyme activity.
\newblock {\em Nature Chemistry}, 5(12):1011--1018, 2013.

\bibitem{kauffman1993origins}
Stuart~A. Kauffman.
\newblock {\em The Origins of Order: {{Self}}-Organization and Selection in
  Evolution}.
\newblock {Oxford University Press}, 1993.

\bibitem{lancet1994emergence}
Doron Lancet, Ora Kedem, and Yizhaq Pilpel.
\newblock Emergence of order in small autocatalytic sets maintained far from
  equilibrium: {{Application}} of a probabilistic receptor affinity
  distribution ({{RAD}}) model.
\newblock {\em Berichte der Bunsengesellschaft f{\"u}r physikalische Chemie},
  98(9):1166--1169, 1994.

\bibitem{vasas2012evolution}
Vera Vasas, Chrisantha Fernando, Mauro Santos, Stuart Kauffman, and E{\"o}rs
  Szathm{\'a}ry.
\newblock Evolution before genes.
\newblock {\em Biology Direct}, 7(1):1, 2012.

\bibitem{hordijk2014conditions}
Wim Hordijk and Mike Steel.
\newblock Conditions for evolvability of autocatalytic sets: {{A}} formal
  example and analysis.
\newblock {\em Origins of Life and Evolution of Biospheres}, 44(2):111--124,
  2014.

\bibitem{orgel2004prebiotic}
Leslie~E. Orgel.
\newblock Prebiotic chemistry and the origin of the {{RNA}} world.
\newblock {\em Critical Reviews in Biochemistry and Molecular Biology},
  39(2):99--123, 2004.

\bibitem{higgs2016effect}
Paul~G. Higgs.
\newblock The effect of limited diffusion and wet\textendash{}dry cycling on
  reversible polymerization reactions: Implications for prebiotic synthesis of
  nucleic acids.
\newblock {\em Life}, 6(2):24, 2016.

\bibitem{ross2016dry}
David~S. Ross and David Deamer.
\newblock Dry/wet cycling and the thermodynamics and kinetics of prebiotic
  polymer synthesis.
\newblock {\em Life}, 6(3):28, 2016.

\bibitem{hargrave2018computational}
Mason Hargrave, Spencer~K. Thompson, and David Deamer.
\newblock Computational models of polymer synthesis driven by
  dehydration/rehydration cycles: Repurination in simulated hydrothermal
  fields.
\newblock {\em Journal of Molecular Evolution}, 86(8):501--510, 2018.

\bibitem{rajamani2008lipidassisted}
Sudha Rajamani, Alexander Vlassov, Seico Benner, Amy Coombs, Felix Olasagasti,
  and David Deamer.
\newblock Lipid-assisted synthesis of {{RNA}}-like polymers from
  mononucleotides.
\newblock {\em Origins of Life and Evolution of Biospheres}, 38(1):57--74,
  2008.

\bibitem{dasilva2015saltpromoted}
Laura Da~Silva, Marie-Christine Maurel, and David Deamer.
\newblock Salt-promoted synthesis of {{RNA}}-like molecules in simulated
  hydrothermal conditions.
\newblock {\em Journal of Molecular Evolution}, 80(2):86--97, 2015.

\bibitem{deguzman2014generation}
Veronica DeGuzman, Wenonah Vercoutere, Hossein Shenasa, and David Deamer.
\newblock Generation of oligonucleotides under hydrothermal conditions by
  non-enzymatic polymerization.
\newblock {\em Journal of Molecular Evolution}, 78(5):251--262, 2014.

\bibitem{flory1953principles}
Paul~J. Flory.
\newblock {\em Principles of Polymer Chemistry}.
\newblock {Cornell University Press}, 1953.

\bibitem{gupta1987reaction}
Santosh~K. Gupta and Anil Kumar.
\newblock {\em Reaction Engineering of Step Growth Polymerization.}
\newblock Plenum Chemical Engineering Series. {Plenum Press}, 1987.

\bibitem{voet2011fundamentals}
Donald Voet and Judith~G. Voet.
\newblock {\em Fundamentals of Biochemistry}.
\newblock {Wiley}, fourth edition, 2011.

\bibitem{gao2019synthesis}
Hongbing Gao, Xiaodong Ma, Jiaping Lin, Liquan Wang, Chunhua Cai, Liangshun
  Zhang, and Xiaohui Tian.
\newblock Synthesis of nanowires via temperature-induced supramolecular
  step-growth polymerization.
\newblock {\em Macromolecules}, 52(20):7731--7739, 2019.

\bibitem{nelson2013lehninger}
David~L. Nelson and Michael~M. Cox.
\newblock {\em Lehninger Principles of Biochemistry}.
\newblock {W. H. Freeman and Company}, {New York}, sixth edition, 2013.

\bibitem{nam2017abiotic}
Inho Nam, Jae~Kyoo Lee, Hong~Gil Nam, and Richard~N. Zare.
\newblock Abiotic production of sugar phosphates and uridine ribonucleoside in
  aqueous microdroplets.
\newblock {\em Proceedings of the National Academy of Sciences},
  114(47):12396--12400, 2017.

\bibitem{orgel1998polymerization}
L.~E. Orgel.
\newblock Polymerization on the rocks: {{Theoretical}} introduction.
\newblock {\em Origins of Life and Evolution of the Biosphere}, 28(3):227--234,
  1998.

\bibitem{monnard2003eutectic}
Pierre-Alain Monnard, Anastassia Kanavarioti, and David~W. Deamer.
\newblock Eutectic phase polymerization of activated ribonucleotide mixtures
  yields quasi-equimolar incorporation of purine and pyrimidine nucleobases.
\newblock {\em Journal of the American Chemical Society}, 125(45):13734--13740,
  2003.

\bibitem{ellis2001macromolecular}
R.~John Ellis.
\newblock Macromolecular crowding: Obvious but underappreciated.
\newblock {\em Trends in Biochemical Sciences}, 26(10):597--604, 2001.

\bibitem{costanzo2009generation}
Giovanna Costanzo, Samanta Pino, Fabiana Ciciriello, and Ernesto~Di Mauro.
\newblock Generation of long {{RNA}} chains in water.
\newblock {\em Journal of Biological Chemistry}, 284(48):33206--33216, 2009.

\bibitem{flory1944thermodynamics}
Paul~J. Flory.
\newblock Thermodynamics of heterogeneous polymers and their solutions.
\newblock {\em The Journal of Chemical Physics}, 12(11):425--438, 1944.

\bibitem{luo2017polymerizationlike}
Binbin Luo, John~W. Smith, Zixuan Wu, Juyeong Kim, Zihao Ou, and Qian Chen.
\newblock Polymerization-like co-assembly of silver nanoplates and patchy
  spheres.
\newblock {\em ACS Nano}, 11(8):7626--7633, 2017.

\bibitem{yang2018supramolecular}
Chaoying Yang, Xiaodong Ma, Jiaping Lin, Liquan Wang, Yingqing Lu, Liangshun
  Zhang, Chunhua Cai, and Liang Gao.
\newblock Supramolecular ``step polymerization'' of preassembled micelles: A
  study of ``polymerization'' kinetics.
\newblock {\em Macromolecular Rapid Communications}, 39(5):1700701, 2018.

\bibitem{gu2019reversible}
Mengxin Gu, Xiaodong Ma, Liangshun Zhang, and Jiaping Lin.
\newblock Reversible polymerization-like kinetics for programmable
  self-assembly of {{DNA}}-encoded nanoparticles with limited valence.
\newblock {\em Journal of the American Chemical Society}, 141(41):16408--16415,
  2019.

\bibitem{oivanen1998kinetics}
Mikko Oivanen, Satu Kuusela, and Harri L{\"o}nnberg.
\newblock Kinetics and mechanisms for the cleavage and isomerization of the
  phosphodiester bonds of {{RNA}} by {{Br{\o}nsted}} acids and bases.
\newblock {\em Chemical Reviews}, 98(3):961--990, 1998.

\bibitem{rackauckas2017differentialequations}
Christopher Rackauckas and Qing Nie.
\newblock {{DifferentialEquations}}.jl \textendash{} a performant and
  feature-rich ecosystem for solving differential equations in {{Julia}}.
\newblock {\em Journal of Open Research Software}, 5(1):15, 2017.

\bibitem{oosawa1975thermodynamics}
Fumio Oosawa and Sho Asakura.
\newblock {\em Thermodynamics of the Polymerization of Protein}.
\newblock Molecular {{Biology}}. {Academic Press}, {London}, 1975.

\bibitem{flory1936molecular}
Paul~J. Flory.
\newblock Molecular size distribution in linear condensation polymers.
\newblock {\em Journal of the American Chemical Society}, 58(10):1877--1885,
  1936.

\bibitem{lu2019supramolecular}
Yingqing Lu, Liang Gao, Jiaping Lin, Liquan Wang, Liangshun Zhang, and Chunhua
  Cai.
\newblock Supramolecular step-growth polymerization kinetics of pre-assembled
  triblock copolymer micelles.
\newblock {\em Polymer Chemistry}, 10(25):3461--3468, 2019.

\bibitem{xing2019indepth}
Ji-Yuan Xing, Yao-Hong Xue, Zhong-Yuan Lu, and Hong Liu.
\newblock In-depth analysis of supramolecular interfacial polymerization via a
  computer simulation strategy.
\newblock {\em Macromolecules}, 52(17):6393--6404, 2019.

\end{thebibliography}

\end{document}